\documentclass[twocolumn,prl,showpacs]{revtex4}
\usepackage{amsmath}
\usepackage{graphicx}
\usepackage{verbatim}
\usepackage[dvipdfm]{epsfig}
\begin{document}

\title{Four wave mixing oscillation in a semiconductor microcavity: Generation of two correlated polariton populations}

\author{M. Romanelli, C. Leyder, J.Ph. Karr, E. Giacobino, and A. Bramati}
\affiliation{Laboratoire Kastler Brossel, Universit\'{e} Paris 6,
Ecole Normale Sup\'{e}rieure et CNRS,\\
UPMC Case 74, 4 place Jussieu, 75252 Paris Cedex 05, France}
\date{\today}
\begin{abstract}
We demonstrate a novel kind of polariton four wave mixing
oscillation. Two pump polaritons scatter towards final states that
emit two beams of equal intensity, separated both spatially and in
polarization with respect to the pumps. The measurement of the
intensity fluctuations of the emitted light demonstrates that the
final states are strongly correlated.
\end{abstract}


\maketitle

In strong-coupling semiconductor microcavities~\cite{weisbuch} the
system is described in terms of "mixed" exciton-photon
quasiparticles, the cavity polaritons. These composite bosons have
very interesting properties~\cite{issue}. Due to their photon
component, polaritons with a given transverse wave vector
$\mathbf{k}$ can be directly created by an incident laser beam with
the appropriate energy and momentum. For the same reason, the
angular distribution of the emission provides information about the
polariton population along the dispersion curve~\cite{houdré94}. The
exciton part is responsible for the coupling between polariton modes
via the Coulomb interaction, which is at the origin of polaritonic
nonlinearities. The understanding of these nonlinearities was
greatly improved by the results of Savvidis et al.~\cite{savvidis},
who demonstrated parametric amplification of a probe beam at normal
incidence, when a polariton population was created at a specific
wave vector $\mathbf{k_{p}}$ by a resonant pump beam. Detection of a
weak "idler" beam with wave vector $\mathbf{2k_{p}}$ allowed to
identify the nonlinear mechanism as parametric scattering of two
polaritons from the pump mode into a signal-idler pair
$\{\mathbf{k_{p}}, \mathbf{k_{p}}\} \Rightarrow \{\mathbf{0},
\mathbf{2k_{p}}\}$~\cite{ciuti}. The system has many similarities
with an optical parametric oscillator (OPO)~\cite{baumberg}, further
underlined by the demonstration of parametric
oscillation~\cite{stevenson}, phase coherence of the nonlinear
emission~\cite{messin} and more recently optical
bistability~\cite{bistamag}. A remarkable property of OPOs above the
oscillation threshold is the generation of two bright quantum
correlated beams~\cite{heidmann}. In principle, polariton parametric
oscillation should lead to the same effect and generate two
macroscopically populated polariton modes (the signal and idler
polaritons) which are quantum correlated~\cite{karr, schwendimann}.
In practice, however, a major problem arises because of the
asymmetry between the signal and idler modes. Due to its large wave
vector, the idler has a small photonic fraction, typically $\simeq 5
\%$. Moreover, its energy is very close to the bare exciton energy,
and this allows a very efficient relaxation towards large
$\mathbf{k}$ excitonic states. As a result, the idler has a large
linewidth compared to the signal, and a much weaker intensity (the
intensity ratio is typically $\sim 10^2-10^4$~\cite{stevenson, saba,
butté}). This makes it very difficult to measure the correlations of
the signal and idler above the oscillation threshold. Even though a
signature of quantum pair correlations has been recently shown in
the regime of multimode parametric scattering below the oscillation
threshold~\cite{Savasta:05}, to the best of our knowledge a direct
evidence of signal-idler correlations in the oscillation regime is
lacking.

\begin{figure}[b]
\epsfig{file=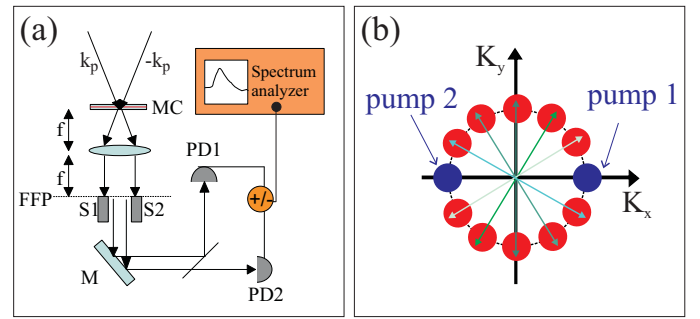,width=0.5\textwidth} \caption{(a)
Experimental setup. Two pump beams $\mathbf{k_{p}}$ and
$\mathbf{-k_{p}}$ are focused on the microcavity sample MC. The
microcavity emission is collimated by a lens (f = 50 mm). Two
movable screens S1 and S2 placed in the far field plane (FFP)
eliminate the transmitted pump beams. The upper part of the emission
is detected by photodiode PD1, the lower part by photodiode PD2. The
photocurrent sum and difference are then spectrally analyzed. (b)
Representation in $\mathbf{k}$-space of the expected scattering
processes. Correlated pairs are connected by arrows.}
\label{fig:setup}
\end{figure}

In this paper, we describe a polariton four wave mixing oscillator
that generates equilibrated signal and idler beams. Above the
oscillation threshold, these beams have the same intensity, since
they are produced by polariton states having opposite wave vector,
and hence the same photonic fraction and linewidth. Furthermore,
they are spatially separated and linearly polarized, with orthogonal
polarization with respect to the pumps: These features greatly
simplify the optical detection, since they permit to filter out the
pump light and the secondary emission produced by elastic scattering
(resonant Rayleigh scattering). We have measured the intensity
correlations of the two beams, finding that almost perfect
correlations are present. In the strong coupling regime, extracavity
photon fields carry the amplitude and phase information of the
intracavity polariton fields; therefore, our results demonstrate the
generation of two strongly correlated macroscopic polariton
populations, in which the electronic excitations in the
semiconductor quantum well are involved.

The experimental setup is shown in fig.~\ref{fig:setup}(a). We use
two pump beams resonant with the lower polariton branch, having
opposite in-plane wave vectors $\{\mathbf{k_{p}},
-\mathbf{k_{p}}\}$. For crossed scattering processes involving one
polariton from each pump mode, momentum conservation requires that
the signal and idler have opposite wave vectors, while energy
conservation imposes $|\mathbf{k}| = |\mathbf{k_{p}}|$. As a
consequence, all pair scattering processes $\{\mathbf{k_{p}},
-\mathbf{k_{p}}\}\Rightarrow \{ \mathbf{k}, -\mathbf{k} \}$ with the
condition $|\mathbf{k}| = |\mathbf{k_{p}}|$ are in principle
allowed.
\begin{figure}[b]
\epsfig{file=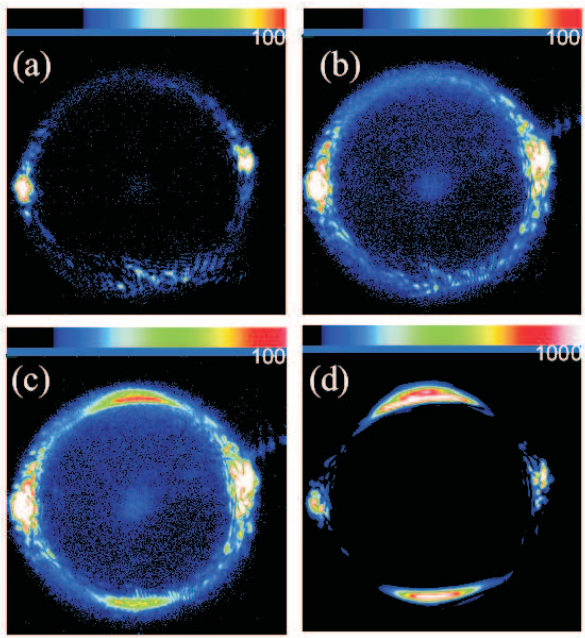,width=0.45\textwidth} \caption{(a) Far
field emission in the TE-polarization for a total pump power of: (a)
3 mW; (b) 19 mW; (c) 27 mW; (d) 39 mW. Cavity-exciton detuning
$\delta$ = 0. At low pump power (a-b), light intensity is roughly
uniform on the elastic circle (except for the two pump spots, lying
approximately on the horizontal diameter). At high pump power (c-d),
scattered polaritons accumulate around the vertical diameter.}
\label{fig:farfield}
\end{figure}
Nonlinear emission is expected on a circle in the far field plane,
whose diameter is fixed by the pump wave vector (see
fig.~\ref{fig:setup}(b)). The incidence angle of the pump beams is
of about 6$^\circ$. This value is chosen so as to avoid competition
with the one-pump scattering channel $\{\mathbf{k_{p}},
\mathbf{k_{p}}\}\Rightarrow \{\mathbf{0}, \mathbf{2k_{p}}\}$ the
efficiency of which is maximal at the so-called "magic angle" i.e.
about 12$^\circ$ for this sample. The two linearly TM-polarized
(polarization in the incidence plane $\leftrightarrow$) pumps are
focused on the microcavity sample on a spot of 80 $\mu$m diameter.
The polarization resolved far field of the transmitted microcavity
emission is obtained using a 50mm lens and imaged on a CCD camera.
Two independently movable screens allow to select a part of the far
field emission, which is subsequently detected by two identical
photodiodes. The two photocurrents are amplified and their sum or
difference is sent to a spectrum analyzer. The microcavity sample is
cooled at 4K in a cold finger cryostat. The sample is described in
detail in Ref.\cite{houdre3}. It is a high quality factor 2$\lambda$
GaAs/AlAs cavity, with three low indium content
In$_{0.04}$Ga$_{0.96}$As quantum wells, one at each antinode of the
cavity mode. The Rabi splitting energy is 5.1 meV. Polariton
linewidths are in the 100-200 $\mu$eV range. The laser source is a
cw Ti:Sa laser, intensity stabilized, with 1 MHz linewidth.

In fig.\ref{fig:farfield} we show the far field emission detected on
the TE-linear polarization, \emph{orthogonal} to the pump
polarization, for several values of the pump power. For low pump
power (a-b), the emitted TE-polarized light is weak and roughly
uniformly distributed on the elastic circle. For higher pump power
(c-d), two bright spots located on the vertical diameter appear,
breaking the circular symmetry of the emission. This surprising
feature is a strong indication that the oscillation threshold has
been reached. Above threshold, polaritons do not redistribute
uniformly over the elastic circle, rather they scatter
preferentially towards states localized around a well defined
diameter, the vertical one. As a result, two macroscopic polariton
populations build up in these states (fig.\ref{fig:farfield}(c-d)).
We note that related phenomena of polarization inversion in the
stimulated scattering regime have been recently
observed~\cite{Kavokina, Renucci, Dasbach}. Our results can be
explained by taking into account the polariton spin
dynamics~\cite{Shelykh, Kavokinb}; this issue will be discussed in
detail in a forthcoming paper.

\begin{figure}[b]
\epsfig{file=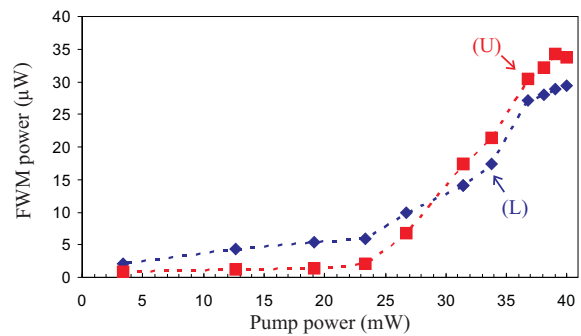,width=0.45\textwidth} \caption{Optical
power of the upper (U) and lower (L) beam as a function of the pump
power. Cavity-exciton detuning $\delta$ = 0.} \label{fig:seuil}
\end{figure}

In fig.~\ref{fig:seuil}, we show the optical power emitted by the
upper and lower polariton populations as a function of the pump
power. Oscillation threshold is reached at about 20 mW of pump
power. Everywhere in text, "pump power" means the sum of the power
of each pump beam. Above threshold, the power ratio of the two beams
is equal to about 1.15. We stress that this situation is very
different from the conventional "magic-angle" configuration, where
the signal beam is $10^2-10^4$ times stronger than the
idler~\cite{stevenson,saba, butté}.

In the case of four-wave mixing emission, the two conjugated beams
are expected to exhibit extremely strong intensity correlations,
since every scattering process from the two pumps must create one
polariton in each of the states on the vertical diameter, in order
to fulfill energy-momentum conservation. We have verified this
feature by measuring the intensity correlations of the two beams. In
order to do so, we have detected the beams with two identical
photodiodes. The generated photocurrents are electronically added or
subtracted; the noise spectra of the photocurrent sum and difference
are measured with a spectrum analyzer, as shown in
Fig.~\ref{fig:setup}.
\begin{figure}[b]
\epsfig{file=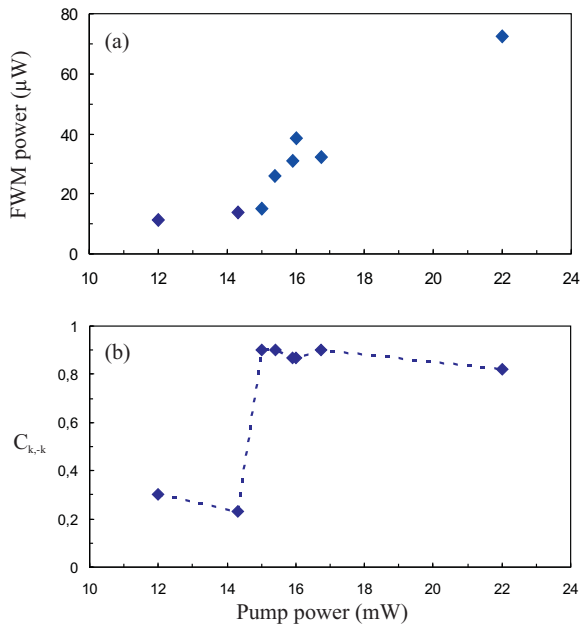,width=0.45\textwidth} \caption{(a) Four
wave mixing optical power as a function of the pump power. (b)
Normalized intensity correlation $C_{\mathbf{k},\mathbf{-k}}$ as a
function of the pump power.} \label{fig:correl}
\end{figure}
In all the experiments described here, the frequency of analysis is
fixed at the same value of 4 MHz. We did not observe any frequency
dependence in the experimentally available frequency range (2-30
MHz). This is due to the fact that the width of the correlation
spectrum is expected to be of the same order as the width of the
polariton resonance, i.e. a few tens of GHz. The normalized
intensity correlation between the beams $\{ \mathbf{k}, -\mathbf{k}
\}$ is defined as

\begin{equation}
C_{\mathbf{k},\mathbf{-k}}(\Omega) =
\frac{S_{\mathbf{k},\mathbf{-k}}
(\Omega)}{\sqrt{S_{\mathbf{k}}(\Omega) S_{\mathbf{-k}}(\Omega)}}
\end{equation}

where $S_{\mathbf{k},\mathbf{-k}}$ is the intensity correlation
spectrum (defined as the Fourier transform of the correlation
function $C_{\mathbf{k},\mathbf{-k}}(\tau) = \langle \delta
I_{\mathbf{k}} (t) \delta I_{\mathbf{-k}} (t+\tau) \rangle$) and
$S_{\mathbf{k}},S_{\mathbf{-k}}$ are the intensity noise spectra of
the two beams (defined as the Fourier transform of the intensity
autocorrelation function $C_{I}(\tau) =  \langle \delta I (t) \delta
I(t+\tau) \rangle$). This quantity verifies $\left|
C_{\mathbf{k},\mathbf{-k}}(\Omega) \right| \leq 1$ and is equal to 1
for perfect correlations and -1 for perfect anticorrelations.

One can have access to the normalized intensity correlation
$C_{\mathbf{k},\mathbf{-k}}$ by measuring the noise spectra of the
intensity sum $S_{+}$, of the intensity difference $S_{-}$, and of
each beam $S_{\mathbf{\pm k}}$. In fact, these quantities satisfy
the following equation:

\begin{equation}
C_{\mathbf{k},\mathbf{-k}} = (S_{+} - S_{-})/4 \sqrt{S_{\mathbf{k}}
S_{\mathbf{-k}}}
\end{equation}

The normalized intensity correlation is plotted as a function of
pump power in Fig.~\ref{fig:correl}(b). One recognizes a clear
threshold behaviour of the correlation coefficient: for low pump
powers the correlation is weak, while it jumps abruptly at values
close to 1 at $\sim$ 15 mW of pump power. Above this value, the
correlation does not exhibit a marked dependence on the pump power.
By comparing Fig.~\ref{fig:correl}(b) with the emitted average power
for this series of experimental data (Fig.~\ref{fig:correl}(a)), one
can see that the correlation increases abruptly at the four wave
mixing oscillation threshold.
\begin{figure}[b]\epsfig{file=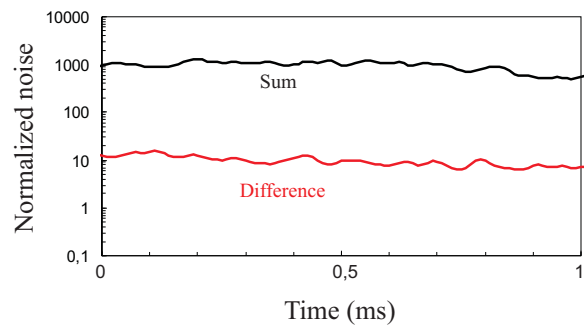,width=0.45\textwidth}
\caption{Noise power of the photocurrent sum and of the photocurrent
difference. The curves are taken at a fixed noise frequency of 4 MHz
and normalized to the standard quantum limit.} \label{fig:noise}
\end{figure}
Such a high value of $C_{\mathbf{k},\mathbf{-k}}$ proves that
polariton four wave mixing allows the generation of two macroscopic
polariton populations that are strongly correlated. Note that the
correlations cannot in any way be due to the fact that the two pump
beams are produced by the same laser. In fact, we have carefully
checked that the intensity fluctuations of the two pump beams are
completely uncorrelated at the frequency of analysis of 4 MHz. This
is due to the fact that the noise level of the pump laser is the
standard quantum limit (at a sufficiently high frequency).
Therefore, the pump beams cannot be at the origin of the
correlations between the two emitted beams. Correlations are instead
generated by the four wave mixing process.

In fig.~\ref{fig:noise}, we show an example of the noise spectra of
the sum and difference of the photocurrents. The curves are taken at
a pump power slightly above the oscillation threshold. Since each
individual beam is extremely noisy, a small unbalance can strongly
affect the noise of the photocurrents difference; therefore, we have
slightly attenuated the upper beam in order to compensate for the
power unbalance and minimize the noise of the
difference~\cite{Romanelli:04}. The photocurrent sum is extremely
noisy ($10^3$ the standard quantum limit); on the contrary, the
beams fluctuations almost perfectly cancel out when the
photocurrents are subtracted, and difference noise is $10^2$ times
lower than the sum noise. The corresponding correlation coefficient
is $C_{\mathbf{k},\mathbf{-k}} = 0.98$.

In conclusion, we have demonstrated a novel kind of polariton four
wave mixing oscillation. In contrast with the so-called magic angle
configuration, our scheme is based on the parametric interaction of
two distinct pump modes with opposite wave vectors. We have shown
that above the oscillation threshold, pump polaritons scatter
towards final states orthogonally polarized with respect to the
pumps, and localized around a well defined diameter of the elastic
circle in the far field. Furthermore, we have demonstrated that the
final states of the scattering process are strongly correlated. We
want here to comment about the fact that, despite the extremely high
value of $C_{\mathbf{k},\mathbf{-k}}$, the correlations are not
\emph{quantum} (since the difference noise is above the standard
quantum limit). We believe that this fact may be interpreted as
follows. As one can see in Fig.~\ref{fig:correl}(b), the normalized
intensity correlation $C_{\mathbf{k},\mathbf{-k}}$ is quite low
below the oscillation threshold, and jumps abruptly at almost 1 just
above threshold. Indeed, this is in contrast with the expected
behavior for a parametric oscillator, which exhibits extremely
strong correlations below threshold. This indicates that the few
$\mu W$ of optical power detected below threshold (see
Fig.~\ref{fig:correl}(a)) are not due to parametric luminescence
below threshold; they come instead from other processes, which give
an uncorrelated background emission, that becomes dominant at low
pump power. Above threshold, this uncorrelated emission may reduce
the observed correlations. If our interpretation is correct, present
results may be improved in a configuration in which the FWM emission
is not exactly at the same energy as the pumps~\cite{Savasta:05},
and therefore do not lie on the elastic circle, where the background
emission is expected to be the strongest.

Our results indicate that the achievement of quantum correlated and
entangled polaritons may be within reach in the near future.

We are very grateful to R. Houdr\'{e} for providing us with the
microcavity sample, and to G.Leuchs for ultralow noise
photodetectors.

\end{document}